\documentclass{article}

\usepackage[english]{babel}

\usepackage[letterpaper,top=2cm,bottom=2cm,left=3cm,right=3cm,marginparwidth=1.75cm]{geometry}
\usepackage[utf8]{inputenc} 
\usepackage[T1]{fontenc}    
\usepackage{hyperref}       
\usepackage{url}            
\usepackage{booktabs}       
\usepackage{amsfonts}       
\usepackage{nicefrac}       
\usepackage{microtype}      
\usepackage{lipsum}
\usepackage{fancyhdr}       
\usepackage{graphicx}       
\graphicspath{{media/}}     

\usepackage{latexsym}
\usepackage{graphicx}
\usepackage{mathptmx}
\usepackage[T1]{fontenc}
\usepackage{lipsum}
\usepackage{algorithm}
\usepackage{algpseudocode}
\usepackage{soul}
\usepackage{rotating} 

\algnewcommand\algorithmicinput{\textbf{Input:}}
\algnewcommand\Input{\item[\hspace{0.65cm}\algorithmicinput]}

\usepackage{amsmath}
\usepackage{amsfonts}
\usepackage{amssymb}
\usepackage{amsbsy}
\usepackage{amsthm}

\usepackage{float}
\usepackage{booktabs}
\usepackage{tabularx}
\usepackage{bm}

\newtheorem{theorem}{Theorem}

\newtheorem{remark}{Remark}

\newtheorem{lemma}{Lemma}

\newcommand{\bb}[1]{\mathbb{#1}}
\renewcommand{\v}[1]{\bm{#1}}
\newcommand{\m}[1]{\mathbf{#1}}

\newcommand{\reals}{\mathbb{R}}
\newcommand{\lt}{\left}
\newcommand{\rt}{\right}

\DeclareMathOperator*{\minimize}{minimize\,}

\DeclareMathOperator*{\subjectto}{subject\,\, to\,}
\DeclareMathOperator*{\diag}{Diag\,}
\def\boxit#1{\vbox{\hrule\hbox{\vrule\kern6pt
          \vbox{\kern6pt#1\kern6pt}\kern6pt\vrule}\hrule}}

\def\boxit#1{\vbox{\hrule\hbox{\vrule\kern6pt
          \vbox{\kern6pt#1\kern6pt}\kern6pt\vrule}\hrule}}

\newcolumntype{C}{>{\centering\arraybackslash}X}

\usepackage{authblk}

\title{Group COMBSS: Group Selection via Continuous Optimization}

\author[1]{Anant Mathur\thanks{Corresponding author: \texttt{anant.mathur@unsw.edu.au}}}
\author[1]{Sarat Moka}
\author[2,3]{Benoit Liquet}
\author[1]{Zdravko Botev}

\affil[1]{School of Mathematics and Statistics, University of New South Wales, NSW, AUSTRALIA}
\affil[2]{School of Mathematical and Physical Sciences, Macquarie University, NSW, AUSTRALIA}
\affil[3]{Laboratoire de Math\'ematiques et de leurs Applications, Universit\'e de Pau et des Pays de l'Adour, Pau, FRANCE}

\begin{document}

\maketitle

\begin{abstract}
We present a new optimization method for the group selection problem in linear regression. In this problem, predictors are assumed to have a natural group structure and the goal is to select a small set of groups that best fits the response. The incorporation of group structure in a predictor matrix is a key factor in obtaining better estimators and identifying associations between response and predictors. Such a discrete constrained problem is well-known to be hard, particularly in high-dimensional settings where the number of predictors is much larger than the number of observations.  We propose to tackle this problem by framing the underlying discrete binary constrained problem into an unconstrained continuous optimization problem.  
The performance of our proposed approach is compared to state-of-the-art variable selection strategies on simulated data sets. 
We illustrate the effectiveness of our approach on a genetic dataset to identify  grouping of markers across chromosomes.
\end{abstract}

\section{INTRODUCTION}
\label{sec:intro}
Given a dataset $(\v y,\,  \m X)$ consisting of a response vector $\v y \in \reals^{n}$ and a design matrix $\m X \in \reals^{n \times p}$ with $n$ and $p$ denoting the number of observations and the number of features respectively, the linear regression assumes that $\v y$ and $\m X$ have the linear relationship,
\begin{align}
\v y = \m X \v \beta + \v \epsilon,     
\label{eqn:linear-reg}
\end{align}
where $\v \beta \in \reals^p$ denotes the unknown regression coefficients and $\v \epsilon = (\epsilon_1, \dots, \epsilon_n)^\top \in \reals^n$ represents a vector of unknown errors, unless otherwise specified, assumed to be independent and identically distributed.

The goal of group selection methods is to identify which groups of features are relevant for predicting the outcome variable and estimate the corresponding regression coefficients. This can help in situations where predictor variables naturally fall into meaningful groups or where there is prior knowledge suggesting that certain groups of variables may be related to the outcome variable.
For instance, in genomics, genes belonging to the same pathway typically share similar functionalities and collaborate in regulating biological systems. The collective effect of these genes can be significant, making it feasible to detect them as a group, either at the pathway or gene set level. Incorporating this grouping structure has become increasingly common, largely due to the success of geneset enrichment analysis approaches \cite{subramanian2005gene}. Incorporating group structure into regression analysis has proven effective for biomarker identification \cite{yuan2006model,meier2008group,puig2009multidimensional,simon2012standardization}. 


To formulate this problem, partition the design matrix $\m X$ into distinct groups, denoted as $\m X = [\m X_{1}, \m X_{2}, \ldots,\m X_{J}]$, where each $\m X_{j} \in \reals^{n \times p_j}$ represents $j$-th group with $p_j$ features. Note that $p = p_1 + \cdots + p_J$. 
Then, \eqref{eqn:linear-reg} can be re-expressed as
\begin{equation}
\label{eqnointercept}
\v y = \sum_{j=1}^{J} \m X_{j} \v \beta_j + \v \epsilon, 
\end{equation}
where 
for each $j$, $\v \beta_j \in \reals^{p_j}$ is the regression coefficients associated with $j$-th group $\m X_{j}$. To simplify our exposition, we do not include an intercept term in \eqref{eqnointercept}, assuming that the response variable $\v y$ is centered.

The group selection problem then can be stated as a subset selection problem of the form
\begin{align}
\minimize_{\v \beta_1,\dots,\v \beta_J} \frac{1}{n}\|\v y - \sum_{j=1}^J\m X_j\v \beta_j\|_2^2,\quad \subjectto \sum_{j=1}^J I\left(\|\v \beta_j\|_2 > 0\right) \leq k.
\label{eqn:group-selection}
\end{align}
where $k$ is the sparsity parameter, $\| \cdot\|_2$ denotes $\mathcal L_2$-norm, and $I(\cdot)$ denotes the usual indicator function. 

By incorporating group-wise structure into the regression model, group selection methods can improve model interpretability, reduce overfitting, and provide insights into the relationships between different groups of features and the outcome variable. 
Common approaches for group selection in linear regression include {\em group Lasso Regression} \cite{yuan2006model}, a variant of the Lasso regression where the penalty term is applied at the group level rather than at the individual variable level thus encouraging sparsity at the group level, effectively selecting entire groups of features. An extension of group Lasso is {\em sparse group Lasso} \cite{simon2013sparse} which allows for within-group sparsity, meaning not all features within a group are forced to be nonzero simultaneously. A third variant is {\em hierarchical variable selection}, which can be useful when the groups exhibit a hierarchical organization, such as in gene expression data or nested experimental designs.  Relatively recent work \cite{hazimeh2023grouped} proposes an efficient approximate algorithm for solving \eqref{eqn:group-selection} based on a combination of coordinate descent and local search methods.

The paper is organized as follows. In Section~\ref{sec:group-combss}, we state the group selection problem and formulate our continuous extension. In Section~\ref{sec:num sim}, we provide extensive numerical experiments comparing the proposed method with the most popular existing methods. In Section \ref{real-data} we demonstrate our method using a complex genetic dataset where single nucleotide polymorphisms (SNPs) are utilized to predict gene expression across four distinct tissue types. Concluding remarks and possible future research directions are in Section~\ref{sec:conclusion}.
 
\section{Group Selection via COMBSS}
\label{sec:group-combss}
The goal of this section is to show how  the (non-group) model selection approach in \cite{moka2024combss} can be extended to the case of  group selection. To this end, we first restate the exact group selection problem \eqref{eqn:group-selection} as a binary constrained problem given by
\begin{align}
\minimize_{s_1,\dots, s_J \in \{0, 1\}} \minimize_{\v \beta_1,\dots,\v \beta_J} \frac{1}{n}\|\v y - \sum_{j=1}^J s_j \m X_j\v \beta_j\|_2^2,\quad \subjectto \sum_{j=1}^J s_j \leq k.
\label{eqn:group-selection-bc}
\end{align}





For each $J$-dimensional binary vector $\v s = (s_1, \dots, s_J) \in \{0,1\}^J$, let $\m X_{[\v s]}$ be matrix constructed from $\m X$ by removing groups $\m X_j$ that correspond to all $s_j = 0$. Thus, the number of columns of $\m X_{[\v s]}$ is equal to $\sum_{j = 1}^J p_j I(s_j  = 1)$. Similarly, let $\v \beta_{[\v s]}$ be the vector obtained from $\v \beta$ by removing the elements of $\v \beta$ indices that correspond all groups with $s_j = 0$. Then, \eqref{eqn:group-selection-bc} can be expressed as 
\begin{align}
\minimize_{\v s \in \{0,1\}^J}\minimize_{\v \beta_{[\v s]}} \frac{1}{n}\| \v y - \m X_{[\v s]}\v \beta_{[\v s]}\|_2^2,\quad \subjectto |\v s| \leq k,
\label{eqn:group-selection-bc2}
\end{align}
where $|\v s|$ denotes the number of 1's in $\v s$.
Now suppose, for a given $\v s$, $\widehat{\v \beta}_{[\v s]}$ is a solution of 
\begin{align}
\minimize_{\v \beta_{[\v s]}} \frac{1}{n}\| \v y - \m X_{[\v s]}\v \beta_{[\v s]}\|_2^2,
\label{eqn:fixedS-minimize}
\end{align}
then \eqref{eqn:group-selection-bc2} is equal to
\begin{align}
\minimize_{\v s \in \{0,1\}^J} \frac{1}{n}\| \v y - \m X_{[\v s]} \widehat{\v \beta}_{[\v s]}\|_2^2,\quad \subjectto |\v s| \leq k.
\label{exp:minimize-for-S}
\end{align}
Solving \eqref{eqn:fixedS-minimize} for $\widehat{\v \beta}_{[\v s]}$ is relatively easier task compared to solving \eqref{exp:minimize-for-S}. Indeed, the latter problem is well-known to be NP-hard \cite{natarajan1995sparse}.

Now, for each $\v t = [t_1, \dots, t_J]^\top \in [0, 1]^J$, let 
\[
\m T_{\v t} = \diag\Big([\underbrace{t_1, \dots, t_1}_{p_1\,\, \text{times}}, \underbrace{t_2, \dots, t_2}_{p_2\,\, \text{times}}, \dots, \underbrace{t_J, \dots, t_J}_{p_J\,\, \text{times}}]^\top\Big)
\]
where $\diag(\v u)$ is a diagonal matrix with diagonal being $\v u$.
Furthermore, take 
\[
\m X_{\v t} = \m X\m T_{\v t},
\]
and define
\begin{align}
\label{eqn:Lt}
\m L_{\v t} = \frac{\m X_{\v t}^\top \m X_{\v t}}{n} +  ( \m I - \m T_{\v t}^2).
\end{align}
Let $\widetilde{\v \beta}_{\v t}$ be a solution of the linear equation (in $\v u$)
\begin{align*}
\m L_{\v t} \v u  = \lt(\frac{\m X_{\v t}^\top \v y}{n} \rt).
\end{align*}
 Then, we consider a Boolean relaxation of \eqref{exp:minimize-for-S} given by
\begin{align}
\minimize_{\v t \in [0,1]^J} \frac{1}{n}\| \v y - \m X_{\v t} \widetilde{\v \beta}_{\v t}\|_2^2,\quad \subjectto \sum_{j = 1}^J t_j \leq k.
\label{exp:minimize-for-t}
\end{align}
The following result establishes some key properties of $\widetilde {\v \beta}_{\v t}$ and shows the relationship between \eqref{exp:minimize-for-S} and \eqref{exp:minimize-for-t}. 
\begin{theorem}
\label{thm:theory}
    The following are true.
    \begin{itemize}
        \item[(i)] $\m L_{\v t}$ is non-singular for all $\v t \in (0, 1)^J$.
        \item[(ii)] For any corner point $\v s \in \{0, 1\}^J$,  $\m X_{[\v s]} \widehat{\v \beta}_{[\v s]} = \m X_{\v s} \widetilde{\v \beta}_{\v s}$.
        \item[(iii)] For every sequence of vectors $\v t^{(1)}, \v t^{(2)}, \dots \in (0, 1)^J$ that converges to a point $\v t \in [0, 1]^J$, 
        \[
        \| \v y - \m X_{\v t} \widetilde{\v \beta}_{\v t}\|_2 = \lim_{\ell \to \infty} \| \v y - \m X_{\v t^{(\ell)}} \widetilde{\v \beta}_{\v t^{(\ell)}}\|_2.
        \]
    \end{itemize}
\end{theorem}

The proofs of (i), (ii) and (iii) are  natural extensions of the proofs of Theorem~1, 2 and 3 in \cite{moka2024combss}, and are thus omitted. 

Theorem~\ref{thm:theory} (i) implies that for all interior points $t \in (0, 1)^J$, $\widetilde{\v \beta}_{\v t}$ is unique and is given by $\widetilde{\v \beta}_{\v t} = \m L_{\v t}^{-1} \m X_{\v t}^\top \v y/n $, and (ii) implies that at the corners of the hypercube $[0, 1]^J$,  the value of objective function in \eqref{exp:minimize-for-t} is identical to the value of the objective function in \eqref{exp:minimize-for-S}. Theorem~\ref{thm:theory} (iii) establishes the continuity of the objective function of the Boolean relaxation problem~\eqref{exp:minimize-for-t}.

In this paper, instead of solving \eqref{exp:minimize-for-t}, we consider a relaxation using the Lagrangian form
\[
f_{\lambda}(\v t) = \frac{1}{n}\| \v y - \m X_{\v t} \widetilde{\v \beta}_{\v t}\|_2^2 + \lambda \sum_{j = 1}^J  \sqrt{p_j}t_j,
\]
for a tuning parameter $\lambda > 0$ and aim to solve 
\begin{align}
\minimize_{\v t \in [0,1]^J} f_{\lambda}(\v t).
\label{eqn:box-constrained-problem}
\end{align}
Instead of the sparsity parameter $k$, we now have the parameter $\lambda$  to control the level of the sparsity in the solution. The $\sqrt{p_j}$ term is included to ensure the penalty term is scale-invariant with respect to the group size.
The optimization \eqref{eqn:box-constrained-problem} still has unwieldy box constraints. To get rid of these box constraints, we consider the equivalent unconstrained problem:
\begin{align}
\minimize_{\v w \in \bb{R}^J} g_{\lambda}(\v w),
\label{eqn:unconstrained-problem}
\end{align}
where 
$g_{\lambda}(\v w) = f_{\lambda}(\v t(\v w)), \,\, \v w \in \bb{R}^J,$
with $\v t(\v w) = 1/(1 + \exp(- \v w))$. That is, for each $i =1, \dots, J$, the $j$-th element $t_j$ is obtained by applying the Sigmoid  function on $w_j$. Since the Sigmoid function is  strictly increasing, solving unconstrained problem \eqref{eqn:unconstrained-problem} is equivalent to solving the box-constrained problem~\eqref{eqn:box-constrained-problem}. We use the Adam optimizer, a popular gradient based approach, for solving \eqref{eqn:unconstrained-problem}. See Appendix~\ref{app:derivative} for a derivation of the gradient $\nabla g_{\lambda}$ of the objective function $g_{\lambda}$. Algorithm~\ref{alg:GroupCOMBSS} provides a pseudo-code for the proposed method. It takes the data $(\v y, \m X)$, group sizes $(p_1, \dots, p_J)$, penalty parameter $\lambda$,  initial point $\v w^{(0)}$, and threshold $\tau$ that helps convert the Sigmoid output into a binary one. For the given $\lambda$, $\mathsf{Adam}\lt(\v w^{(0)}, \nabla g_{\lambda}\rt)$ executes the Adam optimizer, which takes $\v w^{(0)}$ as an initial point to provide a solution $\v w$. This $\v w$ is mapped to a point $\v t \in [0, 1]^J$ using the Sigmoid function and then $\v t$ is mapped to a binary vector $\v s \in \{0, 1\}^J$ using the threshold parameter $\tau \in (0, 1)$.

\begin{algorithm}
\caption{Group COMBSS}\label{alg:GroupCOMBSS}
\begin{algorithmic}[1]
  \Input $(\v y, \m X)$, $(p_1, \dots, p_J)$, 
         $\lambda, \v w^{(0)}, \tau$
    \vspace{0.3cm}
  \State $\v w \leftarrow \mathsf{Adam}\lt(\v w^{(0)}, \nabla g_{\lambda}\rt)$
  
  \For{$j = 1$ to $j = p$}
     \State $t_j \leftarrow 1/(1 + \exp(- w_j))$
     \State $s_j \leftarrow \mathbb{I}(t_j > \tau)$ 
  \EndFor   
  
  \State \Return $\m s = (s_1, \dots, s_p)^\top$
\end{algorithmic}
\end{algorithm}

\begin{remark}
\label{rem:ridge}
Recent work \cite{hazimeh2023grouped,mazumder2023subset} suggests that when the signal-to-noise ratio (SNR) is low, additional ridge regularization can improve the prediction performance of the best subset selection. To include such additional ridge penalty in our implementation, we replace $\widetilde{\v \beta}_{\v t}$ with
\begin{align*}
\widetilde{ \v \beta}^{\text{Ridge}}_{\v t} :=\left[ \m X_{\v t}^\top \m X_{\v t} + n (\m I - \m T_{\v t}^2) + 
 \gamma\, \m T_t^2\right]^{-1}\m X_{\v t}^\top \v y.
\end{align*}
The parameter $\gamma$ controls the strength of the ridge penalty. Note that when $\gamma > 0 $ the estimator $\widetilde{ \v \beta}^{\text{Ridge}}_{\v t}$ agrees with the simple ridge estimator at any corner point,
\[
\widetilde{ \v \beta}^{\text{Ridge}}_{\v s} = \left[ \m X_{[\v s]}^\top \m X_{[\v s]} + 
 \gamma\, \m I\right]^{-1} \m X_{[\v s]}^\top \v y, \quad \v s \in \{0, 1\}^J.
\]

\end{remark}

\section{Numerical Simulations}
\label{sec:num sim}
To compare the performance of a variate of group selection methods, we use datasets simulated from  the model:
\begin{equation}
\label{eq: data_gen}
   \v y = \m X \v \beta^* + \v \epsilon,\quad\text{where}\quad\v\epsilon\sim\mathcal{N}(\v 0, \sigma^2 \m I_n),
\end{equation}
where we generate synthetic predictors $\m X = [\m X_{1}, \m X_{2}, \ldots, \m X_{J}]$ with $\m X_j\in \bb{R}^{n\times p_j}$. The predictor matrix $\m X$ is simulated as a multivariate normal with a between-group correlation $\psi$ and within-group correlation $\rho$. We run Group COMBSS Algorithm~\ref{alg:GroupCOMBSS} and compare its statistical performance against the state-of-the-art grouped variable selection methods: L0 Group, Group Lasso, Group MCP and Group SCAD. We implement Group Lasso, Group MCP and Group SCAD with the R package \texttt{grpreg} \cite{breheny2015group}. L0 Group is implemented with the Python software accompanying \cite{hazimeh2023grouped}. To tune the parameter $\lambda$ we generate an independent validation set from the generating process \eqref{eq: data_gen} with identical parameter values for $\rho$ and $\psi$. We then minimize the generalization risk on the validation set over a grid with 100 values. The coefficient $\v \beta^*$ contains $k$ nonzero groups and the nonzero entries of $\v \beta^*$ are all set to 1.

After generating a training and validation set in each simulation, we run Group COMBSS to evaluate the~$\lambda$ that minimizes the generalization risk on the validation set. We denote this minimizer as $\lambda^*$ and the corresponding model coefficient estimate as $\hat{\v \beta}_{\lambda^*}$. The number of correct and incorrect non-zero groups in $\hat{\v \beta}_{\lambda^*}$ are referred to as true positives $(TP)$ and false positives $(FP)$, respectively. Likewise, the number of correct and incorrect zero groups in $\hat{\v \beta}_{\lambda^*}$ are referred to as true negatives $(TN)$ and false negatives $(FN)$, respectively. 
We consider the following performance measures:  
\begin{enumerate}
    \item \textbf{Precision}: Precision is defined as $TP/(TP+FP).$ A precision close to 1 indicates that the method is reliable in its classifications of non-zero groups while minimizing false positives.
    \item \textbf{Recall}: Recall is defined as $TP/(TP+FN).$ A recall close to 1 indicates that the method is reliable in its classifications of non-zero groups while minimizing false negatives.
    \item \textbf{Matthews correlation coefficient (MCC)}:  MCC is defined as, $${\text{MCC}}={\frac {{\mathit {TP}}\times {\mathit {TN}}-{\mathit {FP}}\times {\mathit {FN}}}{\sqrt {({\mathit {TP}}+{\mathit {FP}})({\mathit {TP}}+{\mathit {FN}})({\mathit {TN}}+{\mathit {FP}})({\mathit {TN}}+{\mathit {FN}})}}}.$$ The MCC is a balanced measure that ranges from $-1$ (perfect disagreement) through 0 (no better than random chance) to $+1$ (perfect agreement). 
    \item \textbf{Generalization Risk}: This is defined as $\frac{1}{n}\|\m X\hat{\v \beta}_{\lambda^*}-\m X\v\beta^*\|_2^2.$
\end{enumerate}
We consider the following simulation settings:

\begin{itemize}
    \item Setting 1: $n = 100$, $p = 40$, $\rho = 0.9$, $\psi = 0.2$, $k = 4$ and $p_j = 4$.
    \item Setting 2: $n = 100$, $p = 40$, $\rho = 0.9$, $\psi = 0.5$, $k = 4$ and $p_j = 4$.
    \item Setting 3: $n = 400$, $p = 600$, $\rho = 0.9$, $\psi = 0.2$, $k = 15$ and $p_j = 4$.
    \item Setting 4: $n = 400$, $p = 600$, $\rho = 0.9$, $\psi = 0.5$, $k = 15$ and $p_j = 4$.

\end{itemize}
The value of $\sigma^2$ is chosen to achieve an SNR of either 1 or 3. For each simulation setting, we replicate the simulation 50 times and report the mean value of each performance measure. Standard errors of the mean are provided in parentheses. 

In the low-dimensional and low-group-correlation setting (Setting 1, Table \ref{tbl:1}), Group COMBSS exhibits the highest MCC, Precision, and Recall scores closely followed by L0 Group. Group LASSO, MCP, and SCAD exhibit lower model risk as these are methods that not only select sparse models but also penalize regression coefficients.  However, the performance of these three methods is  inferior compared to Group COMBSS and L0 Group. In these simulations, the ridge penalty $\gamma$ in Group COMBSS is set to zero, thereby excluding any penalization on the regression coefficients. In subsequent simulation efforts, we intend to explore the implications of a non-zero ridge penalty, chosen over a pre-defined grid. In the high-group correlation setting (Setting 2), we observe Group COMBSS achieving the best MCC score. When the signal is strong and group correlation is low, Group COMBSS and L0 group perfectly identify the non-zero groups in the low noise setting (Setting 1, Table \ref{tbl:2}).

\begin{table}[H]
  \begin{center}
  \caption{Low-dimensional, high noise (Setting 1 and 2).}
  \label{tbl:1}
    \begin{tabularx}{\textwidth}{lcCCCC}
        \toprule
        \multicolumn{1}{c}{} & \multicolumn{1}{c}{} & \multicolumn{4}{c}{SNR = 1}  \\
        \cmidrule(rl){3-6} 
          Method         & Setting       & MCC        & Precision             & Recall            & Risk                  \\
        \cmidrule(r){1-1} \cmidrule(l){2-2}\cmidrule(rl){3-6} 

    Group COMBSS & 1           & 0.95 (0.02) & 0.98 (0.01) & 0.95 (0.02) & 17.87 (1.06) \\ 
    L0 Group       &           & 0.91 (0.02) & 0.97 (0.01) & 0.92 (0.02) & 20.41 (1.07) \\ 
	Group LASSO      &         & 0.46 (0.03) & 0.55 (0.01) & 0.99 (0.01) & 16.89 (0.76) \\ 	
	Group MCP      &           & 0.71 (0.03) & 0.74 (0.02) & 0.96 (0.01) &  4.97 (0.20) \\ 
	Group SCAD      &          & 0.59 (0.03) & 0.65 (0.02) & 0.98 (0.01) & 4.85 (0.20) \\ 	
                     &           &              &            &              &          \\ 
    Group COMBSS & 2 	       & 0.74 (0.03) & 0.94 (0.02) & 0.74 (0.02) & 28.62 (1.28) \\ 
    L0 Group       &           & 0.67 (0.03) & 0.92 (0.02) & 0.66 (0.02) & 32.13 (1.18) \\ 
	Group LASSO      &         & 0.41 (0.03) & 0.53 (0.01) & 0.97 (0.01) & 21.43 (0.94) \\ 	
	Group MCP      &           & 0.47 (0.03) & 0.67 (0.03) & 0.74 (0.03) & 7.31 (0.23) \\ 
	Group SCAD      &          & 0.49 (0.04) & 0.62 (0.02) & 0.91 (0.02) & 6.67 (0.3) \\ 	
 \bottomrule
    \end{tabularx}
    \end{center}
\end{table}

\begin{table}[H]
  \begin{center}
  \caption{Low-dimensional, low noise (Setting 1 and 2).}
  \label{tbl:2}
    \begin{tabularx}{\textwidth}{lcCCCC}
        \toprule
        \multicolumn{1}{c}{} & \multicolumn{1}{c}{} & \multicolumn{4}{c}{SNR = 3}  \\
        \cmidrule(rl){3-6} 
          Method         & Setting       & MCC        & Precision             & Recall            & Risk                  \\
        \cmidrule(r){1-1} \cmidrule(l){2-2}\cmidrule(rl){3-6} 

        Group COMBSS & 1 	   & 1.00 (0.00) & 1.00 (0.00) & 1.00 (0.00) & 5.52 (0.27) \\ 
        L0 Group       &           & 1.00 (0.00) & 1.00 (0.00) & 1.00 (0.00) & 5.49 (0.27) \\ 
	Group LASSO       &        & 0.41 (0.03) & 0.52 (0.01) & 0.52 (0.01) & 6.75 (0.31) \\ 
	Group MCP       &          & 0.78 (0.02) & 0.78 (0.02) & 1.00 (0.00) & 1.56 (0.07) \\ 
	Group SCAD       &         & 0.57 (0.03) & 0.63 (0.02) & 1.00 (0.00) & 1.61 (0.08) \\ 
                     &           &              &            &              &          \\ 
        Group COMBSS & 2           & 0.97 (0.01) & 0.98 (0.01) & 0.98 (0.01) & 9.26 (0.55) \\ 
        L0 Group       &           & 0.91 (0.02) & 0.95 (0.02) & 0.96 (0.01) & 10.65 (0.57) \\ 
	Group LASSO       &        & 0.43 (0.03) & 0.53 (0.01) & 1.00 (0.00) & 9.35 (0.44) \\ 
	Group MCP       &          & 0.65 (0.03) & 0.70 (0.02) & 0.95 (0.02) & 3.09 (0.14) \\ 
	Group SCAD       &         & 0.55 (0.04) & 0.62 (0.02) & 0.98 (0.01) & 3.09 (0.13) \\ 
 \bottomrule
    \end{tabularx}
    \end{center}
\end{table}

In the high-dimensional regime, as shown in Tables \ref{tbl:3} and \ref{tbl:4}, Group COMBSS achieves the best group selection among all methods, attaining a Precision score that is significantly closer to 1 in comparison to the Lasso, MCP, and SCAD, which tend to select a higher number of false positives. As discussed in \cite{mazumder2023subset}, it is observed that in cases of high noise (Table \ref{tbl:3}), the subset selection methods (Group COMBSS, L0 Group) yield higher generalization risk scores. Conversely, in the low-noise, low-group correlation scenario (Setting 3, Table \ref{tbl:4}), Group COMBSS exhibits the lowest generalization risk.

\begin{table}[H]
  \begin{center}
  \caption{High-dimensional, high noise (Setting 3 and 4).}
  \label{tbl:3}
    \begin{tabularx}{\textwidth}{lcCCCC}
        \toprule
        \multicolumn{1}{c}{} & \multicolumn{1}{c}{} & \multicolumn{4}{c}{SNR = 1}  \\
        \cmidrule(rl){3-6} 
          Method         & Setting       & MCC        & Precision             & Recall            & Risk                  \\
        \cmidrule(r){1-1} \cmidrule(l){2-2}\cmidrule(rl){3-6} 
        Group COMBSS & 3 & 0.64 (0.01) & 0.80 (0.02) & 0.56 (0.01) & 194.04 (4.89) \\ 
        L0 Group       &           & 0.56 (0.01) & 0.83 (0.02) & 0.42 (0.01) & 238.72 (5.05) \\ 
        Group LASSO    &           &  0.39 (0.01) & 0.28 (0.01) &   0.87 (0.01) & 132.93 (2.82) \\
        Group MCP   &              &  0.49 (0.01) & 0.43 (0.01) & 0.71 (0.02) & 157.16 (3.37)\\
        Group SCAD  &              & 0.41 (0.01) & 0.29 (0.01) & 0.86 (0.01) & 135.59 (2.88)\\
                     &           &              &            &              &          \\
        Group COMBSS & 4 & 0.30 (0.02) & 0.51 (0.02) & 0.23 (0.01) & 313.71 (7.98) \\ 
        L0 Group       &           & 0.25 (0.02) & 0.50 (0.03) &  0.17 (0.01) & 373.94 (7.04) \\ 
        Group LASSO    &           &  0.21 (0.01) & 0.21 (0.01) &  0.55 (0.02) & 171.99 (3.95) \\
        Group MCP   &              &  0.20 (0.01) & 0.28 (0.01) & 0.30 (0.01) & 259.55 (5.99)\\
        Group SCAD  &              & 0.21 (0.01) & 0.21 (0.01) & 0.53 (0.02) & 173.23 (4.30)\\

 \bottomrule
    \end{tabularx}
    \end{center}
\end{table}

\begin{table}[H]
  \begin{center}
  \caption{Low-dimensional, low noise (Setting 3 and 4).}
  \label{tbl:4}
    \begin{tabularx}{\textwidth}{lcCCCC}
        \toprule
        \multicolumn{1}{c}{} & \multicolumn{1}{c}{} & \multicolumn{4}{c}{SNR = 3}  \\
        \cmidrule(rl){3-6} 
          Method         & Setting       & MCC        & Precision             & Recall            & Risk                  \\
        \cmidrule(r){1-1} \cmidrule(l){2-2}\cmidrule(rl){3-6} 
        Group COMBSS & 3 & 0.94 (0.01) & 0.95 (0.01) & 0.94 (0.01) & 55.70 (1.70) \\ 
        L0 Group       &             & 0.88 (0.01) & 0.96 (0.01) &  0.84 (0.01) & 69.27 (2.33) \\ 
        Group LASSO      &           & 0.47 (0.01) & 0.30 (0.00) &  1.00 (0.00) & 58.61 (1.34)\\ 
        Group MCP        &           & 0.73 (0.01) & 0.63 (0.01) & 0.94 (0.01)  &      66.06 (2.05) \\
        Group SCAD      &           & 0.54 (0.01) & 0.36 (0.01)& 0.99 (0.00)  &      64.01 (1.54)   \\
                          &           &              &            &              &          \\ 
        Group COMBSS & 4 & 0.57 (0.02) & 0.74 (0.02) &  0.49 (0.01) & 139.57 (3.44) \\ 
        L0 Group       &           & 0.50 (0.02) & 0.77 (0.02) &  0.37 (0.01) & 172.26 (3.44) \\ 
        Group LASSO       &        & 0.38 (0.01) & 0.27 (0.01) &  0.85 (0.01) & 90.36 (1.95) \\ 
        Group MCP       &        & 0.36 (0.01) & 0.38 (0.01) &  0.49 (0.01) & 153.58 (3.01) \\ 
        Group SCAD       &        & 0.41 (0.01) & 0.30 (0.01) & 0.83 (0.02) &  97.67 (2.63) \\ 

 \bottomrule
    \end{tabularx}
    \end{center}
\end{table}

\section{Illustration with genetic Data} \label{real-data}

We demonstrate the application of our approach within the domain of genetic regulation. In expression Quantitative Trait Loci (eQTL) analysis, aimed at uncovering the genetic factors influencing gene expression variation (i.e., transcription), gene expression data are treated as a quantitative phenotype, while genotype data (SNPs) serve as predictors. In this study, we utilize a dataset extracted from a larger investigation \cite{heinig2010trans} focusing on the Hopx genes, as referenced in \cite{Pedretto}. This dataset has also been analyzed by \cite{liquet2016r2guess}, who employed a Bayesian model to identify a concise set of predictors explaining the collective variability of gene expression across four tissues: adrenal gland (ADR), fat, heart, and kidney. \cite{liquet2017bayesian} utilize a sparse group Bayesian multivariate regression model for a similar objective.   The Hopx dataset comprises 770 SNPs from 29 inbred rats forming the predictor matrix ($n = 29$, $p = 770$), with the expression levels measured in the four tissues (ADR, fat, heart, and kidney) serving as outcomes. A comprehensive description of the dataset is also available in \cite{R2GUESS} and can be accessed through the R package \texttt{R2GUESS}.
Table~\ref{group-appli} displays how the SNPs are distributed across the 20 chromosomes of the rats. The chromosome information establishes the grouping structure of the predictor matrix.


\begin{table}
\centering
\caption{Repartition of the SNPs along the chromosomes.}
\label{group-appli}
\small
\begin{tabular}{|r|@{\hspace{0.3em}}r|@{\hspace{0.3em}}r|@{\hspace{0.3em}}r|@{\hspace{0.3em}}r|@{\hspace{0.3em}}r|@{\hspace{0.3em}}r|@{\hspace{0.3em}}r|@{\hspace{0.3em}}r|@{\hspace{0.3em}}r|@{\hspace{0.3em}}r|@{\hspace{0.3em}}r|@{\hspace{0.3em}}r|@{\hspace{0.3em}}r|@{\hspace{0.3em}}r|@{\hspace{0.3em}}r|@{\hspace{0.3em}}r|@{\hspace{0.3em}}r|@{\hspace{0.3em}}r|@{\hspace{0.3em}}r|
@{\hspace{0.3em}}r|}
\hline 
Chromosome & 1 & 2 & 3 & 4 & 5 & 6 & 7 & 8 & 9 & 10 & 11 & 12 & 13 & 14 & 15 & 16 & 17 & 18 & 19 & 20 \\
\hline
Group size & 74 & 67 & 63 & 60 & 39 & 45 & 52 & 43 & 31 & 51 & 21 & 26 & 33 & 22 & 15 & 27 & 18 & 30 & 34 & 19 \\
\hline
\end{tabular}
\end{table}
\begin{figure}[h]
\centering
\includegraphics[width=12cm, height=6cm]{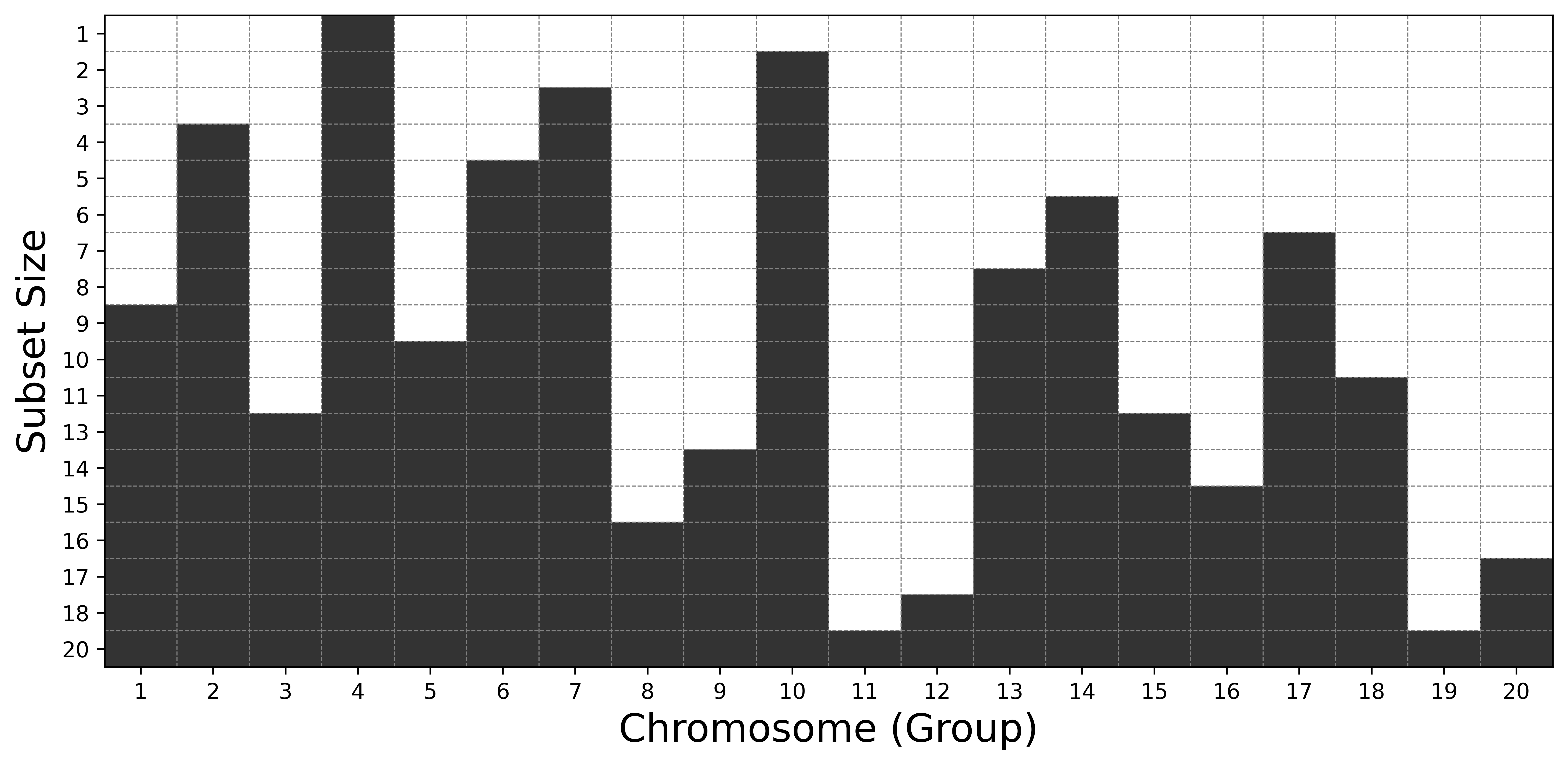}
\caption{Best Subset Solution Path for ADR variable $(\gamma= 1)$. \label{ADR}}
\end{figure}

We executed Group COMBSS on each tissue separately. The best subset solution path for each tissue has been obtained over a grid with 150  values of $\lambda$  using Algorithm~\ref{alg:GroupCOMBSS}. Due to the high-dimensional aspect of the data $(n = 29, p = 770)$, we add a ridge penalization $(\gamma = 1)$. The solution path for the ADR tissue is presented in Figure~\ref{ADR}. As an example, we analyse a parsimonious model with 4 groups, our model picked chromosomes 2, 4, 7 and 10. These chromosomes were also identified as being linked to the ADR tissue  \cite{liquet2017bayesian}, which utilized a sparse group Bayesian multivariate regression model.  The results of our approach on the other tissues are presented in Figures~\ref{fig:kidney}, \ref{fig:heart} and \ref{fig:fat} in Appendix~\ref{app:figures}. For the Kidney tissue, chromosomes 3, 4, 7 and 10 have been selected (for a model with four groups). Note that the ADR and Kidney outcomes are highly correlated ($r=0.7$), which may explain why three out of four groups have common chromosomes. For the Heart Tissue, chromosomes 2, 4, 14 and 15 have been selected (for a model with four groups) while chromosomes 1, 2, 4 and 15 have been selected when analysing the Fat tissue.  Note also that the solution path using COMBSS for a partial least squares approach \cite{liquet2024best}, with a multivariate outcome (the four tissues) but without group selection, has identified a parsimonious set of SNPs located on chromosomes 4, 10, and 14. Chromosome 4 was selected in our four separate models, chromosome 10 was selected with the ADR and Kidney models and finally, chromosome 14 was selected when we analysed the heart tissue.

\section{Conclusion and Future Directions}
\label{sec:conclusion}
In this paper, we presented an unconstrained continuous optimization algorithm for the group selection problem in linear regression. Our approach  makes it possible to extend the non-group selection method in \cite{moka2024combss} to the group selection setting. We conducted extensive numerical simulations in both high- and low-dimensional settings to compare the performance of the proposed algorithm with the popular grouped variable selection approaches.  

We have demonstrated our technique on a complex dataset comprising gene expression data (with four measurements from 29 samples) and SNP explanatory variables (consisting of $770$ variables). The dataset exhibits a structured group organization (with 20 groups), delineated by chromosomes. Our current Group COMBSS selection is not designed yet for analysing a multivariate response.  To fully exploit the multivariate response, one can extend the univariate square error loss to accommodate the multivariate outcome. Furthermore, in genetics, it's a common practice to introduce an additional layer of sparsity within selected groups to improve interpretability. This involves identifying the relevant SNPs (variables) within the chosen groups. 

Sparse group selection problem is an important generalization of the group selection problem, where in addition to the group selection, it is assumed that only a small number of features in each selected group are active. Similar to \cite{friedman2010note} that extends Group LASSO to the sparse group selection problem, our method can be extended to this problem. To see this, in addition to $\v t \in [0, 1]^J$, we consider $\v r = [\v r_1^\top, \dots, \v r_J^\top]^\top \in [0, 1]^p$ with $\v r_j = [r_{j, 1}, \dots, r_{j, p_j}]^\top \in [0, 1]^{p_j}$. The vector $\v r$ acts as binary relaxation for individual features. We can enforce group and within-group sparsity by incorporating appropriate penalties on $\v t$ and $\v r$.

In future research, 
we can also include a ridge penalty as explained in Remark~1 to enhance Group COMBSS's performance when minimizing generalization risk, particularly when SNR is low.

In \cite{moka2024combss}, an alternative version of COMBSS for linear regression is proposed for best subset selection, i.e., optimization \eqref{eqn:group-selection-bc2} with the number of groups equal to the number of features ($J = p$). Future work can focus on numerical and theoretical study of the extension of this version of COMBSS to the group setting. 

\appendix
\section{Derivatives of the objective function}
\label{app:derivative}
Our goal is to solve \eqref{eqn:unconstrained-problem} using a gradient descent approach. To do that, we need to compute the gradient $\nabla_{\v w} g_{\lambda}(\v w) = \left(\partial g(\v w)/\partial w_1, \dots,\partial g(\v w)/\partial w_J\right)$. With $\odot$ denoting the Hadamard (i.e., element-wise) product between two vectors, observe that
\[
\nabla_{\v w} g_{\lambda}(\v w) = \nabla_{\v t} f_{\lambda}(\v t(\v w)) \odot \v t(\v w) (1 - \v t(\v w)),
\]
where we used the fact that the derivative of the Sigmoid function $\v t(\v w)$ is $\v t(\v w)(1 - \v t(\v w))$.

Let $\m Z = \m X^{\top}\m X/n - \m I$, so that $\m L_{\v t} = \m T_{\v t} \m Z  \m T_{\v t} + \m I$. Further, let  $\m E_j$ be a diagonal matrix of dimension $p$ with zeros everywhere except ones along the diagonal at $(\sum_{k = 1}^{j-1} p_k) + 1, \dots, \sum_{k = 1}^{j} p_k$. The following result establishes the derivatives $\partial\widetilde{\v \beta}_{\v t}/\partial t_j$. Its proof is similar to the derivation of the gradient in \cite{moka2024combss} and hence ignored.

\begin{lemma}
Let $\widetilde{\v \beta}_{\v t}  = \m L_{\v t}^{-1}\lt(\frac{\m X_{\v t}^{\top}\v y}{n}\rt)$. For any $\v t \in (0,1)^J$, the derivatives of $\widetilde{\v \beta}_{\v t}$ are given by
\[
\frac{\partial\widetilde{\v \beta}_{\v t}}{\partial t_j} = \m L^{-1}_{\v t}\lt[\m E_j - \m E_j \m Z \m T_{\v t} \m L^{-1}_{\v t}\m T_{\v t}- \m T_{\v t} \m Z \m E_j \m L^{-1}_{\v t} \m T_{\v t}\rt]\lt(\frac{\m X^{\top}\v y}{n}\rt),\quad j=1,\dots,J.
\]
\end{lemma}
We shall use this Lemma to obtain $\nabla f_{\lambda}(\v t)$ for $\v t \in (0,1)^{J}$.  Let $\v \eta_{\v t} =\m T_{\v t} \widetilde{\v \beta}_{\v t}$. Then,
\begin{align*}
    \|\v y - \m X_{\v t} \widetilde{\v \beta}_{\v t}\|_2^2 &= \|\v y - \m X \v \eta_{\v t}\|_2^2
    = \v y^{\top}\v y - 2\v \eta_{\v t}^{\top}\lt(\m X^{\top}\v y\rt)+\v \eta_{\v t}^{\top}\m X^{\top}\m X \v \eta_{\v t}.
\end{align*}
Now we focus on the $j$-th element of $\nabla_{\v t} f_{\lambda}(\v t)$, that is, 
\[
\frac{\partial f_{\lambda}(\v t)}{\partial t_j} =  \frac{\partial }{\partial t_j} \frac{1}{n} \| \v y - \m X_{\v t} \widetilde{\v \beta}_{\v t}\|_2^2 + \sqrt{p_j}\lambda.
\]
Here,
\begin{align*}
    \frac{\partial}{\partial t_j}\lt[\frac{1}{n}\|\v y - \m X_{\v t} \widetilde{\v \beta}_{\v t}\|_2^2\rt] = \frac{2}{n}\lt(\frac{\partial\v \eta_{\v t}}{\partial t_j}\rt)^{\top}\lt[(\m X^{\top}\m X )\v \eta_{\v t}-\m X^{\top}\v y\rt]
    =2\lt(\frac{\partial\v \eta_{\v t}}{\partial t_j}\rt)^{\top}\v a_{\v t},
\end{align*}
where $\v a_{\v t} = \lt(\m X^{\top}\m X/n\rt)\v  \eta_{\v t}-\lt(\m X^{\top}\v y/n\rt)$.
From the definition of  $\widetilde{\v \beta}_{\v t}$ and $\v \eta_{\v t}$,
\begin{align*}
    \frac{\partial \v \eta_{\v t}}{\partial t_j} 
     =\frac{\partial \m T_{\v t}}{\partial t_j} \widetilde{\v \beta}_{\v t}+ \m T_{\v t}\frac{\partial  \widetilde{\v \beta}_{\v t}  }{\partial t_j} 
    = \m E_{j} \widetilde{\v \beta}_{\v t}+ \m T_{\v t} \m L^{-1}_{\v t}\lt[\m E_j -  \m E_j \m Z \m T_{\v t} \m L^{-1}_{\v t} \m T_{\v t}-\m T_{\v t} \m Z  \m E_j \m L^{-1}_{\v t} \m T_{\v t}\rt]\lt(\frac{ \m X^{\top}\v y}{n}\rt).
\end{align*}
Further simplification yields,
\begin{align*}
    \frac{\partial \v \eta_{\v t}}{\partial t_j} &= \m E_{j} \widetilde{\v \beta}_{\v t}+ \m T_{\v t}  \m L^{-1}_{\v t}\lt[ \m E_j\lt(\frac{\m X^{\top}\v y}{n}\rt) - \m E_j \m Z \v \eta_{\v t}- \m T_{\v t} \m Z \m E_j\widetilde{\v \beta}_{\v t}\rt]\\
    &= \m E_{j} \widetilde{\v \beta}_{\v t}-{\m T}_{\v t} \m L^{-1}_{\v t}\m E_j\v b_{\v t} - \m T_{\v t} \m L^{-1}_{\v t}\m T_{\v t}\m Z \m E_j\widetilde{\v \beta}_{\v t},
\end{align*}
where $\v b_{\v t} = \m Z\v \eta_{\v t}-\lt(\frac{\m X^{\top}\v y}{n}\rt)= \v a_{\v t}-\v \eta_{\v t}$.
To further simplify, let $\v c_{\v t} = \m L_{\v t}^{-1}(\v t \odot \v a_{\v t})$, and
$\v d_{\v t} = \m Z(\v t \odot \v c_{\v t})$.
Then, the matrix $\frac{\partial \v \eta_{\v t}}{\partial \v t}$ of dimension $p \times J$ with the $j$-th column being $\frac{\partial \v \eta_{\v t}}{\partial t_j}$ can be expressed as
\[\frac{\partial \v \eta_{\v t}}{\partial \v t}=\operatorname{BlkMat}( \widetilde{\v \beta}_{\v t})-\m T_{\v t}\m L_{\v t}^{-1}\operatorname{BlkMat}(\v b_{\v t})-\m T_{\v t}\m L_{\v t}^{-1}\m T_{\v t}\m Z\operatorname{BlkMat}(\widetilde{\v \beta}_{\v t}),\]
where for a $p$-dimensional vector $\v a_{\v t} = [\v a_{\v t,1}^\top, \dots, \v a_{\v t,J}^\top]^\top$, the $p \times J$ matrix $\operatorname{BlkMat}(\v a_{\v t})$ is defined as
\[
\operatorname{BlkMat}(\v a_{\v t}) := 
\begin{bmatrix} 
    \v a_{\v t,1} & \v 0            & \dots &  \v 0       \\
    \v 0          & \v a_{\v t,2}   &       &  \vdots     \\
    \vdots        & \v 0            & \ddots&  \vdots     \\
    \vdots        & \vdots          &       &  \vdots     \\
    \v 0          & \v 0            & \dots & \v a_{\v t, J}
\end{bmatrix}.
\]
Let $\v h = [\sqrt{p_1},\dots,\sqrt{p_J}]^\top$. Then,
\begin{align*}
    \nabla f_{\lambda}(\v t) &= 2\lt(\frac{\partial\v \eta_{\v t}}{\partial t_j}\rt)^{\top}\v a_{\v t} + \lambda \v h\\
    &= 2\operatorname{BlkMat}(\widetilde{\v \beta}_{\v t})^{\top}\v a_{\v t}-2\operatorname{BlkMat}(\v b_{\v t})^{\top}\m L_{\v t}^{-1}\m T_{\v t}\v a_{\v t}- 2\operatorname{BlkMat}(\widetilde{\v \beta}_{\v t})^{\top}\m Z \m T_{\v t}^{\top}\m L_{\v t}^{-1}\m T_{\v t}\v a_{\v t}+ \lambda \v h\\
    & = 2\operatorname{BlkMat}(\widetilde{\v \beta}_{\v t})^{\top}\v a_{\v t}-2\operatorname{BlkMat}(\v b_{\v t})^{\top}\v c_{\v t}- 2\operatorname{BlkMat}(\widetilde{\v \beta}_{\v t})^{\top}\v d_{\v t}+ \lambda \v h\\
    &= 2\begin{bmatrix} 
    \widetilde{\v \beta}_{\v t,1}^{\top}(\v a_{\v t,1} -  \v d_{\v t,1})-\v b_{\v t,1}^{\top}\v c_{\v t,1}\\
    \vdots         \\
     \widetilde{\v \beta}_{\v t,J}^{\top}(\v a_{\v t,J} -  \v d_{\v t,J})-\v b_{\v t,J}^{\top}\v c_{\v t,J}\\  
\end{bmatrix} + \lambda \v h.
\end{align*}

\newpage
\section{Supplement Material: Genetic Data}
\label{app:figures}

\begin{figure}[h]
\centering
\includegraphics[width=12cm, height=5.3cm]{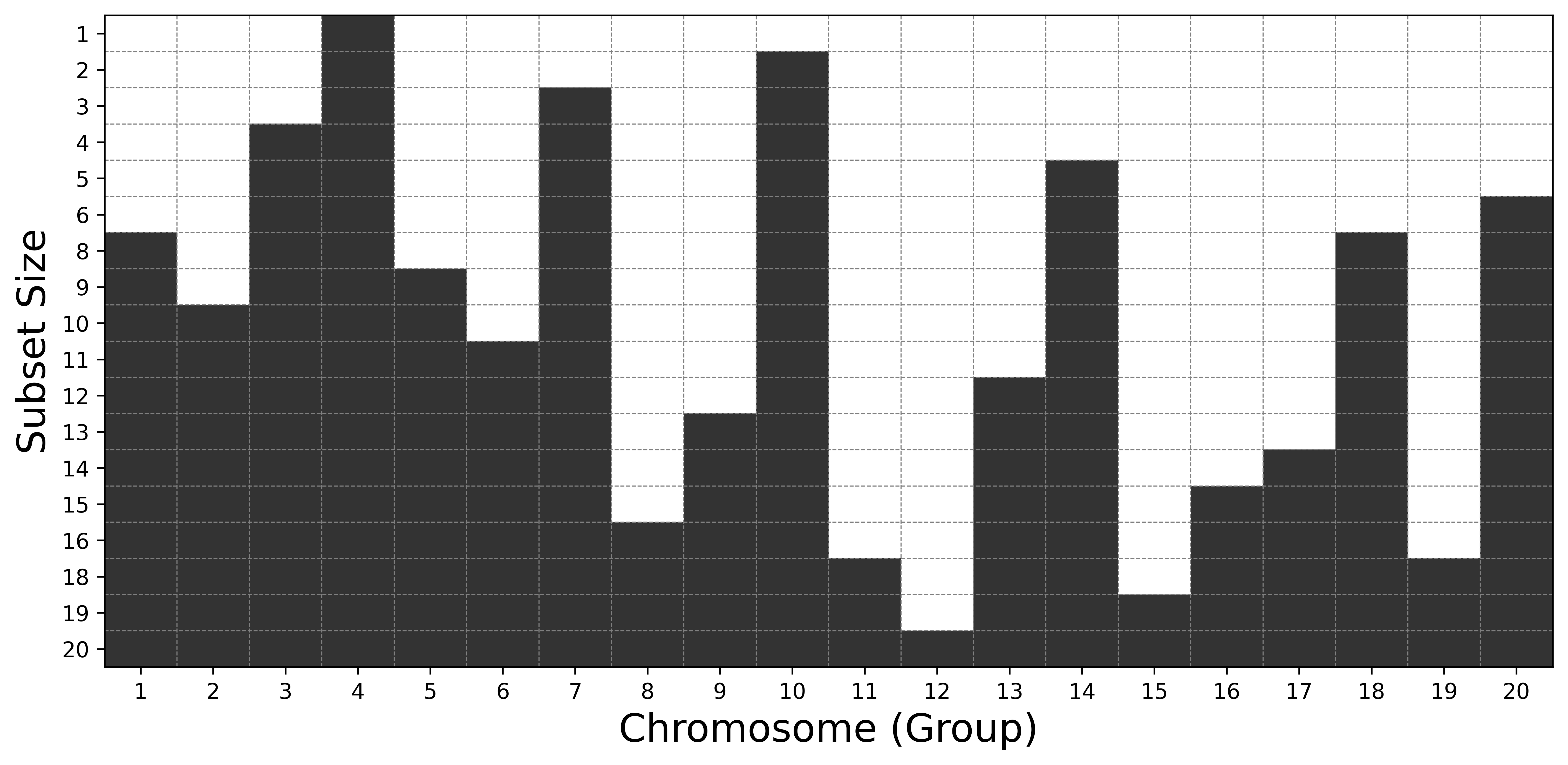}
\caption{Best Subset Solution Path for Kidney variable $(\gamma = 1)$. \label{fig:kidney}}
\end{figure}
\begin{figure}[h]
\centering
\includegraphics[width=12cm, height=5.3cm]{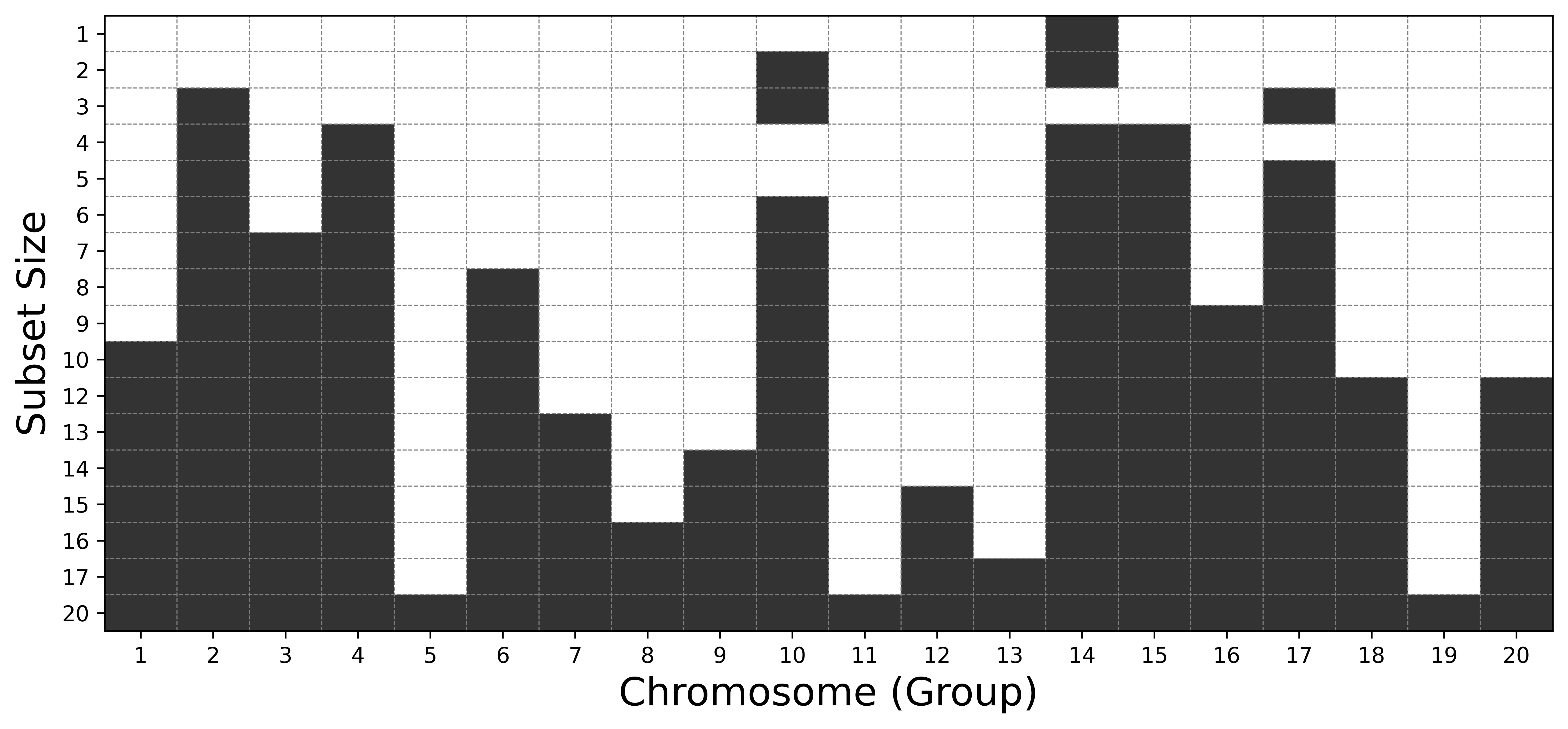}
\caption{Best Subset Solution Path for Heart variable $(\gamma = 1)$. \label{fig:heart}}
\end{figure}
\begin{figure}[h]
\centering
\includegraphics[width=12cm, height=5.3cm]{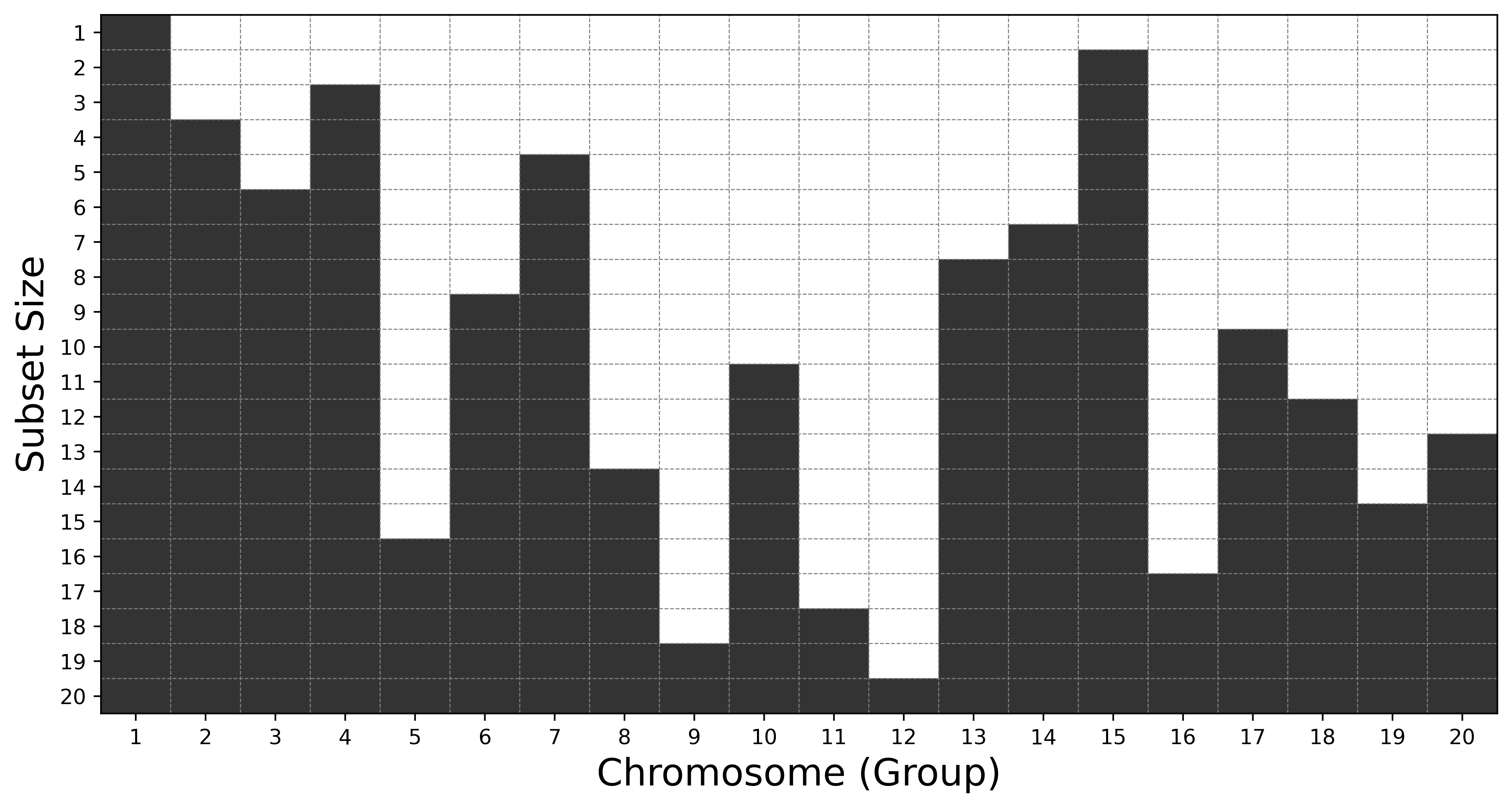}
\caption{Best Subset Solution Path for Fat variable $(\gamma = 1)$. \label{fig:fat}}
\end{figure}

\bibliographystyle{unsrt}  
\bibliography{references}

\end{document}